\documentclass[aip, pop, reprint, onecolumn]{revtex4-1}
\usepackage{natbib}

\usepackage{graphicx}
\usepackage{epstopdf}
\usepackage{amsmath}
\usepackage{amssymb}

\usepackage{color}
\usepackage{xspace}

\newcommand\etc{\textit{etc.}\xspace}
\newcommand\eg{\textit{e.g.}\xspace}
\newcommand\ie{\textit{i.e.}\xspace}

\newcommand{\vpar}{\ensuremath{v_{\parallel}}}
\newcommand{\vperp}{\ensuremath{v_{\perp}}}

\providecommand\bnabla{\boldsymbol{\nabla}}
\providecommand\hbnabla{\hat{\boldsymbol{\nabla}}}

\newcommand{\nzoL}[1]{{\cal L}\left[#1\right]}

\def \z {\hat{\bf z}}

\def \kperp {k_{\perp}}
\def \Lpar {L_{\parallel}}
\def \vth {v_{\mathrm{th}}}

\def \vsT {v_*^T}
\def \hvpar {\hat{v}_{\parallel}}
\def \hvperp {\hat{v}_{\perp}}
\def \hz {\hat{z}}

\def \hy {\hat{y}}

\begin{document}

\title{On the scaling of ion and electron temperature gradient driven turbulence in slab geometry}
\author{G. G. Plunk}
\email{gplunk@ipp.mpg.de}
\affiliation{Max Planck Institute for Plasma Physics, Wendelsteinstr. 1, 17491 Greifswald, Germany}

\begin{abstract}
We demonstrate that the scaling properties of slab ion and electron temperature gradient driven turbulence may be derived by dimensional analysis of a drift kinetic system with one kinetic species.  These properties have previously been observed in gyrokinetic simulations of turbulence in magnetic fusion devices.
\end{abstract}

\maketitle

Predicting the properties of turbulence in magnetic fusion plasmas is a theoretical challenge that has a number of complications, such as the requirement of a kinetic description, the presence of curved magnetic geometry, and the interplay of a variety of waves and instabilities.  For this reason, exact results obtained in simple limits have special value.  They can, for instance, help identify the essential physical processes, facilitate the validation of numerical methods for simulating the turbulence, or provide a condition to be satisfied by complicated theories.  One such result is  a scaling law, \ie a relation that expresses the dependence of a turbulence property on basic system parameters (up to a multiplicative factor).  As found in numerical simulations, such laws are actually satisfied in both the tokamak \cite{barnes-cb-itg} and stellarator \cite{plunk-xanthop-prl} contexts, making their importance significantly more than academic.

In this Brief Communication, it is demonstrated that a number of scaling properties of ion temperature gradient (ITG) and electron temperature gradient (ETG) driven turbulence can be derived by simple symmetry arguments in slab geometry.  This derivation, following the method of \citet{connor-taylor}, reveals a rigorous basis for the existence of these laws, with minimal assumptions.

To begin, let us denote the length of the turbulence domain in the $z$ direction (parallel to the mean magnetic field, ${\bf B}_0 = \z B_0$) as $\Lpar$.  In the perpendicular $x$-$y$ directions, the domain can be assumed sufficiently large to contain many correlation lengths of the turbulence, but much smaller than the scale of variation of the background plasma.  The $x$-$y$ domain therefore does not introduce any further characteristic scales to the problem, and we can assume triply periodic boundary conditions for definiteness.

We consider one kinetic species, which could be either ions or electrons, and the other species is assumed to be adiabatic.  We will thus drop the species labels in what follows.  The collisionless gyrokinetic equation, driven only by a temperature gradient, can be reduced in the limit $\kperp \rho \ll 1$ to the drift kinetic system

\begin{equation}
\frac{\partial f}{\partial t} + \frac{\z\times\bnabla\phi}{B_0}\cdot\bnabla f + \vpar \frac{\partial f}{\partial z} = \left(\vsT \frac{\partial}{\partial y} - \vpar \frac{\partial}{\partial z}\right) \frac{q\phi}{T_0} F_0,\label{drift-kinetic-eqn}
\end{equation}

\noindent where the perturbed distribution function is $f({\bf r}, \vperp, \vpar, t)$ where ${\bf r} = \hat{\bf x} x +\hat{\bf y} y + \z z$, we employ the following other definitions: $n_0$ and $T_0$ are the background density and temperature, $\vsT = \rho\vth/L_T [v^2/(2\vth^2) - 3/2]$, $v^2 = \vperp^2 + \vpar^2$, $\rho = m \vth/(q B_0)$, $q$ is the charge (note that $q_i = Ze$ and $q_e = -e$), $\vth = \sqrt{T/m}$, $L_T^{-1} = d\ln T_0/dx$, and $F_0 = n_0(2\pi\vth^2)^{-3/2}\exp[-v^2/(2\vth^2)]$ is the background Maxwellian distribution.  The electrostatic potential $\phi$ is determined by the quasi-neutrality constraint, which can be written as

\begin{equation}
 \frac{q\tau}{T_0}\nzoL{\phi} = \frac{2\pi}{n_0}\int_{-\infty}^{\infty} d\vpar \int_{0}^{\infty}\vperp d\vperp f,\label{qn-eqn}
\end{equation}

\noindent where $\tau = T_{0i}/ZT_{0e}$ for ITG and $\tau = Z T_{0e}/T_{0i}$ for ETG, where $Z = q_i/e$.  For ITG, we assume a modified Boltzmann response, where the flux surface average in slab geometry is taken to be an average over the $y$-$z$ plane:

\begin{equation}
\nzoL{\phi} = \begin{cases}
\phi - \frac{1}{L_y \Lpar}\int dz dy \phi \quad &\mathrm{for\ ITG,}\\
\phi  \quad &\mathrm{for\ ETG.}
\end{cases}
\end{equation}

\noindent Note that for ITG this implies zero density for the ``zonal'' component, $k_y = k_z = 0$, and therefore to for that case only, we must also include the zonal vorticity equation (obtained from the gyrokinetic system, including terms of order $k^2 \rho^2$; see \citet{plunk-njp} Eqn. D.9), which is expressed succinctly in Fourier space, \eg $\phi = \sum_{\bf k} \phi_{\bf k} \exp(i {\bf k}\cdot{\bf r})$, as

\begin{equation}
k_x^2 \frac{\partial \phi_{\bf k}}{\partial t} = \sum_{{\bf p},{\bf q}} \frac{\epsilon({\bf k}, {\bf p}, {\bf q})}{B_0} \phi_{\bf p}^* \left[q^2 \phi_{\bf q}^* - 2{\bf p}\cdot{\bf q}\chi_{\bf q}^*\right],\quad k_y = k_z = 0.\label{z-vorticity-eqn}
\end{equation}

\noindent where $\epsilon({\bf k}, {\bf p}, {\bf q}) = \delta({\bf k} + {\bf p} + {\bf q})\hat{\bf z}\cdot({\bf p}\times{\bf q})$ and we define $\chi = 2\pi/(q n_0) \int \vperp d\vperp d\vpar (m \vperp^2/4)f$. 

Following \citet{connor-taylor}, the scaling behavior of turbulence can be deduced by performing scaling transformations on the fundamental governing equations -- by this approach, it was proved that electrostatic gyrokinetic turbulence must exhibit gyro-Bohm transport scaling \cite{hagan-frieman}.  In the present case, we apply the following nondimensionalization:

\begin{align}
\hat{t}  &= \frac{t\vth}{\Lpar}    &\hvperp &= \frac{\vperp}{\vth}        &\hvpar  &= \frac{\vpar}{\vth}\nonumber\\
\hat{z} &= \frac{z}{\Lpar}         &\hat{x} &= \frac{x L_T}{\rho \Lpar}          &\hat{y}                &= \frac{y L_T}{\rho \Lpar}\label{nondim-eqn}\\
\hat{f}  &= \frac{\vth^3 L_T^2}{n_0 \rho \Lpar}f      &\hat{\phi} &= \frac{q L_T^2}{T_0 \rho \Lpar}\phi          &\hat{\chi} &= \frac{q L_T^2}{T_0 \rho \Lpar}\chi\nonumber
\end{align}

\noindent The resulting dimensionless equations are

\begin{eqnarray}
\tau \nzoL{\hat{\phi}} = 2\pi \int_{-\infty}^{\infty} d\hvpar \int_{0}^{\infty}\hvperp d\hvperp \hat{f},\label{qn-n-eqn}\\
\frac{\partial \hat{f}}{\partial \hat{t}} + \z\times\hbnabla\hat{\phi}\cdot\hbnabla \hat{f} + \hvpar \frac{\partial \hat{f}}{\partial \hz} = \left( [\hat{v}^2/2 - 3/2]\frac{\partial}{\partial \hy} - \hvpar \frac{\partial}{\partial \hz}\right) \hat{\phi} \frac{\exp[-\hat{v}^2/2]}{(2\pi)^{3/2}},\label{drift-kinetic-n-eqn}\\
\hat{k}_x^2 \frac{\partial \hat{\phi}_{\bf \hat{k}}}{\partial \hat{t}} = \sum_{{\bf \hat{p}},{\bf \hat{q}}} \epsilon({\bf \hat{k}},{\bf \hat{p}},{\bf \hat{q}}) \hat{\phi}_{\bf \hat{p}}^* \left[\hat{q}^2 \hat{\phi}_{\bf \hat{q}}^* - 2{\bf \hat{p}}\cdot{\bf \hat{q}} \hat{\chi}_{\bf \hat{q}}^*\right],\quad \hat{k}_y = \hat{k}_z = 0,\label{z-vorticity-n-eqn}
\end{eqnarray}

\noindent where $\hbnabla$ is the gradient in the normalized variables, and we note the distinction between the normalized coordinates $\hz$, \etc, and the (bold) basis vectors $\z$, {\em etc}.  The only remaining system parameter in the normalized system, which appears only in Eqn.~\ref{qn-n-eqn}, is the temperature ratio $\tau$, defined above, and therefore the properties of the turbulence as measured in the normalized variables, \eg characteristic amplitudes and scales, \etc, can depend only on $\tau$.  Consequently, the characteristic scales in the $x$ and $y$ direction, although they need not be the same, must scale as $\rho \Lpar/L_T$, and the characteristic time must scale as $\Lpar/\vth$, and so forth, in accordance with the transformation \ref{nondim-eqn}.  Furthermore, writing the heat flux in terms of the dimensionless quantities, we obtain the scaling law

\begin{equation}
Q =  \vth n_0 T_0 \frac{\rho^2 \Lpar}{L_T^3}\hat{Q},
\end{equation}

\noindent where the dimensionless quantity $\hat{Q} = 2\pi \sum_{\bf \hat{k}} i \hat{k}_y \hat{\phi}_{\bf \hat{k}} \int \hvperp d\hvperp d\hvpar \hat{f}_{\hat{\bf k}} \hat{v}^2/2$ can only depend on $\tau$, as mentioned.  This is the scaling obtained by \citet{barnes-cb-itg} for ITG turbulence in the tokamak context.  We have derived it, however, for both ETG and ITG turbulence in the slab context, while avoiding assumptions such as isotropy in the $x$-$y$ plane, and other conjectures.  Note that in either the ETG or ITG case, the system is inherently anisotropic (in the $x$-$y$ plane) due to the drive ($\vsT$) term, so it is significant that $x$-$y$ isotropy is not needed to derive the scaling laws.  For the case of ITG, in particular, we can conclude that the characteristic length scale associated with the zonal component must have the same scaling as that of the non-zonal part of the turbulence, though the turbulence need not be strictly isotropic -- \ie there can be some ($\tau$-dependent) asymmetry to the $x$-$y$ spectrum.

It should be noted that the above derivation relied on the presence of only a single drive, the temperature gradient, and the absence of a (stabilizing) density gradient -- the drift kinetic system therefore has no critical temperature gradient below which the ITG mode is stable.  In the presence of a finite density gradient, the scaling should thus only be expected to apply for temperature gradients well above the critical one.  Nevertheless, the approach applied here might also be applied to other cases where a single drive mechanism is dominant, \eg the parallel velocity gradient, or the density gradient, to obtain similar scaling laws.  This could be useful for studying turbulence in different magnetic confinement devices, where the presence (or absence) of such ideal scaling behavior could give a hint about the relative importance of the underlying instabilities and related saturation mechanisms.

\bibliography{slab-itg-etg-scaling}

\end{document}